\documentclass[debug, preprint, twocolumn]{rmaa}
\usepackage{natbib}
\usepackage{graphicx} 
\usepackage{amsmath}
\usepackage{hyperref}
\usepackage{listings}
\usepackage{titlesec}
\usepackage{xcolor}
\hypersetup{
    colorlinks,
    linkcolor={black!50!black},
    citecolor={blue!50!black},
    urlcolor={blue!80!black}
}
\titleformat{\section}{\normalfont\large\bfseries\centering}{\thesection}{1em}{\MakeUppercase}
\usepackage{perpage}
\usepackage{wrapfig}
\MakePerPage{footnote}
\usepackage{newunicodechar}
\lstdefinestyle{nrgstyle}{
  basicstyle=\ttfamily\small,
  commentstyle=\color{gray},
  showstringspaces=false,
  breaklines=true,
  frame=single,
  captionpos=b
}
\usepackage{paralist}
\usepackage{caption}  
\usepackage{psfrag,color}

\usepackage[utf8]{inputenc}

{\catcode`\$=\active
}%

\newcommand{\Cloudy}{\textsc{Cloudy}}
\newcommand{\Hazy}{\textsc{Hazy}}
\newcommand{\cdCommand}[1]{\textbf{#1}}

\font\manual=manfnt at 7pt \def\dbend{\hbox{\raise0.9ex\hbox{\manual\char127\hspace{0.6em}}}}

\newcounter{INTERNALionstage}

\def\gtsim{\mathrel{\hbox{\rlap{\hbox{\lower4pt\hbox{$\sim$}}}\hbox{$>$}}}}
\def\lesssim{\mathrel{\hbox{\rlap{\hbox{\lower4pt\hbox{$\sim$}}}\hbox{$<$}}}}

\def\h0{\mbox{{\rm H}$^0$}}
\DeclareMathAlphabet{\vib}{OML}{cmm}{m}{it}

\usepackage{common/aas_macros}

\title{The 2025 release of \Cloudy}

\author{Chamani M. Gunasekera\altaffilmark{1},
Peter A. M. van Hoof\altaffilmark{2},
Maryam Dehghanian\altaffilmark{3},
Priyanka Chakraborty\altaffilmark{4,5},
Gargi Shaw\altaffilmark{6},
Stefano Bianchi\altaffilmark{7},
Marios Chatzikos\altaffilmark{3}, 
Masahiro Tsujimoto\altaffilmark{8},
Gary J. Ferland\altaffilmark{3}
}

\altaffiltext{1}{Space Telescope Science Institute, Baltimore, MD, USA}
\altaffiltext{2}{Royal Observatory of Belgium, Ringlaan 3, B-1180 Brussels, Belgium}
\altaffiltext{3}{Physics \& Astronomy, University of Kentucky, Lexington, Kentucky, USA}
\altaffiltext{4}{Center for Astrophysics $\vert$ Harvard \& Smithsonian, Cambridge, MA, USA}
\altaffiltext{5}{University of Arkansas, Physics and Astronomy, 825 W Dickson St, Fayetteville, Arkansas 72701}
\altaffiltext{6}{Department of Astronomy and Astrophysics, Tata Institute of Fundamental Research, Mumbai 400005, India}
\altaffiltext{7}{Dipartimento di Matematica e Fisica, Universit\`a degli Studi Roma Tre, via della Vasca Navale 84, 00146 Roma, Italy}
\altaffiltext{8}{Institute of Space and Astronautical Science (ISAS), Japan Aerospace Exploration Agency (JAXA),
3-1-1 Yoshinodai, Chuo-ku, Sagamihara, Kanagawa 252-5210, Japan}

\fulladdresses{
 \item C.{} Gunasekera, Physics \& Astronomy, University of Kentucky, Lexington KY 40506 (cmgunasekera@uky.edu).
 \item P.{} A.{} M.{} van Hoof, Royal Observatory of Belgium, Ringlaan 3, 1180 Brussels, Belgium (p.vanhoof@oma.be).
 \item M.{} Dehghanian, Physics \& Astronomy, University of Kentucky, Lexington KY 40506 (m.dehghanian@uky.edu).
 \item P.{} Chakraborty, Center for Astrophysics $\vert$ Harvard \& Smithsonian, Cambridge, MA (priyanka.chakraborty@cfa.harvard.edu).
 \item G.{} Shaw, Department of Astronomy and Astrophysics, Tata Institute of Fundamental Research, Mumbai 400005, India (gargishaw@gmail.com).
 \item S.{} Bianchi, Dipartimento di Matematica e Fisica, Universit\`a degli Studi Roma Tre, via della Vasca Navale 84, 00146 Roma, Italy (stefano.bianchi@uniroma3.it).
 \item M.{} Chatzikos, Physics \& Astronomy, University of Kentucky, Lexington KY 40506 (mchatzikos@uky.edu).
 \item M.{} Tsujimoto, Institute of Space and Astronautical Science (ISAS), Japan Aerospace Exploration Agency (JAXA), 3-1-1 Yoshinodai, Chuo-ku, Sagamihara, Kanagawa 252-5210, Japan (tsujimoto.masahiro@jaxa.jp).
 \item G.{} Ferland, Physics \& Astronomy, University of Kentucky, Lexington KY 40506 (gary@uky.edu). 
 }

\shortauthor{Gunasekera et al}
\shorttitle{Cloudy's 2025 Release}

\listofauthors{M. Gunasekera et al}
\indexauthor{Gunasekera, C.M.}
\indexauthor{Ferland, G.J.}
\indexauthor{Guzm\'an, F.}
\indexauthor{Jonathan S. Milby}
\indexauthor{Shaw, G.}
\indexauthor{Sarkar, Arnab}
\indexauthor{Chakraborty, Priyanka}

\abstract{We present the 2025 release of the spectral synthesis code \Cloudy, highlighting significant enhancements to the scope and accuracy of the physics which have been made since the previous release. A major part of this development involves resolving the Lyman $\alpha$ line into $j$-resolved fine-structure doublets, making \Cloudy{} of use to the X-ray community. On this front, we have also updated inner-shell ionization line energies and incorporated the 1 keV feature commonly observed in X-ray binaries. Additionally, we update our in-house database, Stout, for the carbon isoelectronic sequence, improving \Cloudy{} microphysical calculations for all wavelengths. We have also extended the molecular network by adding new silicon-bearing species, titanium-related reactions, and phosphorus-containing molecules, enhancing \Cloudy’s ability to model the complex chemistry relevant to rapidly growing field of exoplanet atmospheres. Finally, we outline future developments aimed at maximizing the scientific return from the current and upcoming generation of observatories, including XRISM, JWST, Roman, the Habitable Worlds Observatory (HWO) and NewAthena.}
\resumen{
}

\addkeyword{Atomic data}
\addkeyword{Astronomy software}
\addkeyword{Active galaxies}
\addkeyword{Computational methods}
\addkeyword{Galaxy clusters}
\addkeyword{Molecular data}

\begin{document}

\maketitle

\tableofcontents

\section{Introduction}

We introduce the next major release of \Cloudy{}, C25.00. \Cloudy{} is a spectral synthesis code, simulating plasma conditions ranging from highly non-equilibrium conditions to full Local and Strict Thermodynamic Equillibrium (LTE \& STE). 
Beginning from the first principles of physics and chemistry, 
\Cloudy{} self-consistently solves the chemistry, radiation transport, and dynamics to determine the ionization, chemical state, temperature, and excitation of all species. Much of the physics is discussed in \citet{AGN3}. It does so for the full electromagnetic spectrum. 

Ongoing development since 1978 has continually expanded the code's range of capabilities. Table\,\ref{tab:CloudyReleasePapers} lists the previous review papers that capture the state of the code at that time. The last major release was C23.01 \citep{2023RMxAA..59..327C, 2023RNAAS...7..246G}. With each release, we aim to provide users with a tool that maximizes the impact on their research.

A major effort has been undertaken since C23.01, with the goal of enabling maximum science by the full suite of advanced observatories available to us today.
Detailed in Section\,\ref{sec:xray}, the major development in this release was updating \Cloudy{}'s H-like iso-sequence to match the spectral resolution of the X-Ray Imaging and Spectroscopy Mission (XRISM). Additionally, in Section\,\ref{sec:molecular_data} we present updates to our chemistry network to better equip \Cloudy{} for modeling the complex chemistry, characteristic of the rapidly growing field of exoplanet atmospheres. We also present new updates to our atomic data in Section\,\ref{sec:atomic_data}, which allow for more accurate, state-of-the-art microphysical calculations and improving spectral predictions across the full electromagnetic range. 
Lastly, Sections \ref{sec:misc_impovements} \& \ref{sec:infra} introduce new commands and data files designed to enhance usability for \Cloudy{} users. In light of these major developments, we strongly recommend that users upgrade to the latest version of \Cloudy{}, C25.00, which includes bug fixes that enhance the accuracy of synthesized spectra and improve overall usability.

\begin{table*}
\centering
\caption{Major \Cloudy\ release papers}
\label{tab:CloudyReleasePapers}
\begin{tabular}{|c c c|}
\hline
Version & Year & Citation \\
\hline
C$33- 87$ & $1983-1998$ &  
\citet{1991hbic.book.....F,1993hbic.book.....F,1996hbic.book.....F}+\\
C90 & 1998 & \citet{ 1998PASP..110..761F} \\
C13 & 2013 &\citet{2013RMxAA..49..137F}\\
C17 & 2017 & \citet{2017RMxAA..53..385F}\\
C23, C23.01 & 2023 & \citet{2023RMxAA..59..327C}, \citet{2023RNAAS...7..246G}\\ 
\\
\hline
\end{tabular}
\end{table*}


\section{Atomic Data}
\label{sec:atomic_data}

\subsection{Stout}
We have significantly extended the atomic database for the C-like isoelectronic sequence in \Cloudy{} by incorporating a new, comprehensive dataset. The updated dataset includes 590 fine-structure levels per ion, combining high-precision theoretical energies from R-matrix calculations \citep{2025Atoms..13...44D} with experimentally measured energies from the NIST Atomic Spectra Database Version  5.12 \citep{NIST2024}. This expansion includes N~II to Kr~XXXI (i.e., N$^{+}$ to Kr$^{30+}$) and enables a more accurate and complete treatment of excitation, and emission processes in photoionized and collisionally excited plasmas. More detail will be provided in a forthcoming paper, Dehghanian et al., (in preparation).

\subsubsection{Energy levels}
theoretical energies are now enclosed in square brackets (e.g., [12345.6]), to distinguish between experimental and theoretical values in the energy level files. \Cloudy{}'s internal parser has been enhanced to interpret this syntax, automatically flagging such levels as theoretical. This distinction is propagated throughout the simulation, adding a question mark after the wavelength unit where appropriate in the \texttt{.out} file produced by the simulation. Figure~\ref{fig:stout} shows an example from a \Cloudy{} \texttt{.out} file, where the new feature --- uncertain wavelengths derived from theoretical energy levels --- is marked with a question mark.
\begin{figure}[ht]
\captionsetup{type=figure}
\caption{Intrinsic line intensities from a sample \Cloudy{} model}\label{fig:stout}
\begin{lstlisting}[style=nrgstyle]
...
O  3       5005.93A?   -8.714   17.4419
O  3       5006.19A?   -9.581    2.3652
O  3       5006.33A?   -8.964    9.7965
O  3       5006.84A    -5.809 13994.061
O  3       5006.88A?   -2.166 *********
O  3       5006.89A    -6.589 2323.5645
...
\end{lstlisting}
\end{figure}


\subsubsection{Transition probabilities}
We updated the transition probabilities to align with the new energy level structure. Dipole-allowed transitions are drawn from recent 2025 R-matrix calculations \citep{2025Atoms..13...44D}, providing agreement with the corresponding collision strengths across 590 levels per ion. For forbidden lines, we retained the 2020 data \citep{2020A&A...634A...7M}, since they are not available in the 2025 version of R-matrix calculations by \citet{2025Atoms..13...44D}. When experimental energies from NIST were adopted, the associated NIST transition probabilities were used, where available, to maintain consistency. This hybrid approach ensures the TP data are physically self-consistent, combining modern theory with reliable measurements.

\subsubsection{Collisional strengths}

We have also updated \Cloudy{}'s \texttt{.coll} files in the Stout directory for C-like ions using the recent R-matrix calculations from \citet{2025Atoms..13...44D}, which provide improved electron-impact collision strengths across 590 levels per ion. These data agree with the updated energy levels and transition probabilities, ensuring accurate modeling of collisional excitation processes.

\paragraph{Special case of N~II}
For N~II, although we updated the Stout dataset in the same manner as for other ions in the isoelectronic sequence, \Cloudy{} defaults to using the CHIANTI dataset because it is based on a targeted study of singly ionized nitrogen \citep{2011ApJS..195...12T}. This dataset includes 58 energy levels. If the user prefers to use the Stout dataset (which includes 590 levels), they can simply enable it in the file \texttt{data/stout/masterlist/Stout.ini.}

\paragraph{Special case of O~III}
{In our modeling, we adopt a hybrid approach for the [O~III] collision strengths, selecting the most reliable dataset for each set of transitions based on consistency, temperature coverage, and agreement across recent calculations. For transitions among the five lowest levels of O~III 
we use the data from \citet{2014MNRAS.441.3028S} for $T_{\rm e}\leq30,000$~K and then switch to \citet{2025Atoms..13...44D} for higher temperatures. These include the important ground-term fine-structure lines at 88.35~$\mu$m (1--2) and 51.81~$\mu$m (2--3), as well as the optical nebular lines like 5006.84~\AA\ (3--4). 
For all other transitions involving higher excited levels beyond the lowest five, we adopt the more recent data from \citet{2025Atoms..13...44D}. This dataset, which builds upon and corrects the earlier \citet{2020A&A...634A...7M} results, is based on a systematic R-matrix calculation 
across the entire carbon isoelectronic sequence. The Del Zanna dataset resolves previous 
inconsistencies and includes a critical bug fix that impacted the earlier values. While \citet{2025Atoms..13...44D} show strong agreement with  \citet{2014MNRAS.441.3028S} for many transitions, 
noticeable differences remain --- especially for transitions involving level 4, such as 3--4 --- 
primarily at lower temperatures relevant to photoionized plasmas. Given that photoionized clouds can extend to temperatures below 10$^{4}$~K, where such discrepancies significantly impact 
emissivity predictions and derived abundances, our hybrid strategy ensures both consistency with widely used references and incorporation of the most accurate atomic data available for the full temperature range of interest.}

{Figure~\ref{fig:o3cs} compares the temperature-dependent collision strengths (CS) for the three key [O~III] transitions 
using data from three different sources. The red curves show the updated values from \citet{2025Atoms..13...44D}, the blue dashed curves represent the earlier results from \citet{2020A&A...634A...7M}, and the green curves show the values from \citet{2014MNRAS.441.3028S}, which are currently adopted in \Cloudy{} version C23.01. For the 51.81~$\mu$m (2--3) transition, all three datasets are in excellent agreement across the entire temperature range. In the case of the 88.35~$\mu$m (1--2) line, small discrepancies are observed, with the Storey et al. (2014) values slightly lower than the others at low temperatures. However, the most significant deviation appears in the 5006.84~\AA\ (3--4) transition, where both Mao et al. and Del Zanna et al. datasets overestimate the collision strength relative to the Storey et al. dataset, particularly at photoionization temperatures below 10$^{4}$~K. In all three cases, the black dot indicates the temperature at which we switch to \citet{2025Atoms..13...44D} from \citet{2014MNRAS.441.3028S}.
 \begin{figure}[ht]
   \centering
   \includegraphics[width=\columnwidth]{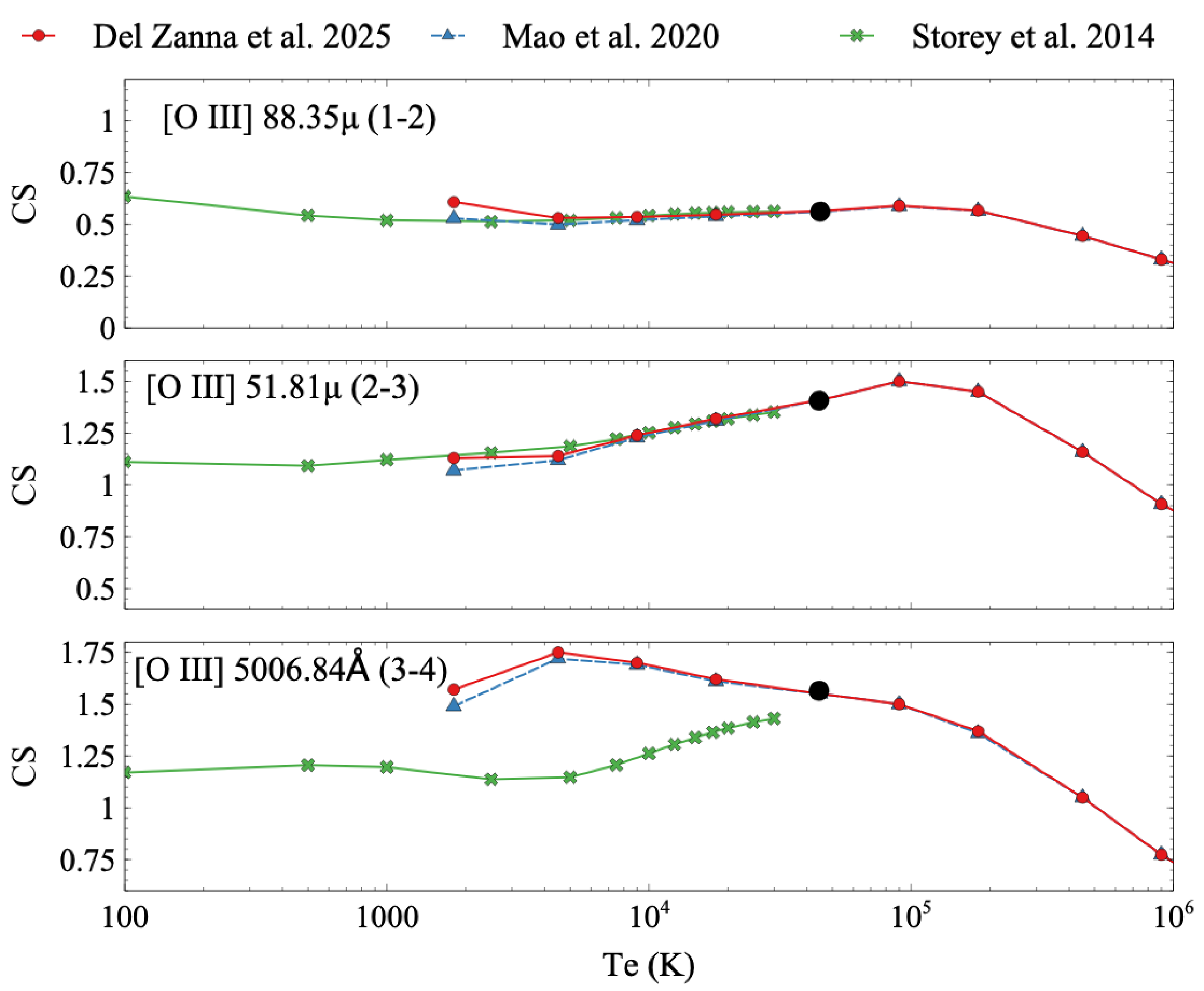}
      \caption{Temperature-dependent thermally-averaged CS for three key [O~III] transitions, showing differences in the 88.35~$\mu$m and 5006.84~\AA\ lines at$T_e<10^4$~K. These discrepancies motivate adopting a hybrid dataset for transitions among the lowest five levels. 
      The black dot on the plots marks this threshold temperature, indicating the point at which we change the reference dataset for the lower levels.}
         \label{fig:o3cs}
   \end{figure}
\paragraph{Special case of Fe~XXI}
While we extended the energy levels to 590 for all C-like isoelectronic ions, we were able to expand the model for Fe~XXI to include 620 levels by adding 30 levels with a K-shell vacancy. For this ion, levels 591-620 and the associated transition probabilities are extracted from \citet{2003A&A...403.1175P}.

The new atomic structure framework improves \Cloudy{}'s predictive power for diagnostic lines in C-like ions across UV and X-ray wavelengths, and supports the demands of modern high-resolution instruments such as JWST and XRISM. We are actively working on extending this framework to additional isoelectronic sequences, with the goal of building a consistently high-fidelity atomic database for use in modern astrophysical modeling. 

To illustrate the impact of the new atomic data on model predictions, Figure~\ref{fig:O3} (taken from Dehghanian et al., in preparation) compares the emission spectrum of O~III (upper panel) and Fe~XXI (lower panel) generated using the previous dataset available in C23.01 (red crosses) and the newly updated dataset (gray dots), based on the C-like model described above. The updated model produces a substantially larger number of emission lines, especially in the infrared, and provides more physically complete predictions. This improvement enables more robust comparisons to high-resolution observations from facilities like JWST and XRISM. The denser distribution of lines also highlights the role of weak transitions that were previously missing or underestimated. Dehghanian et al. (in preparation), will provide a comprehensive review of these updates and details the implementation process.
 \begin{figure*}[t]
   \centering
   \includegraphics[width=0.9\textwidth]{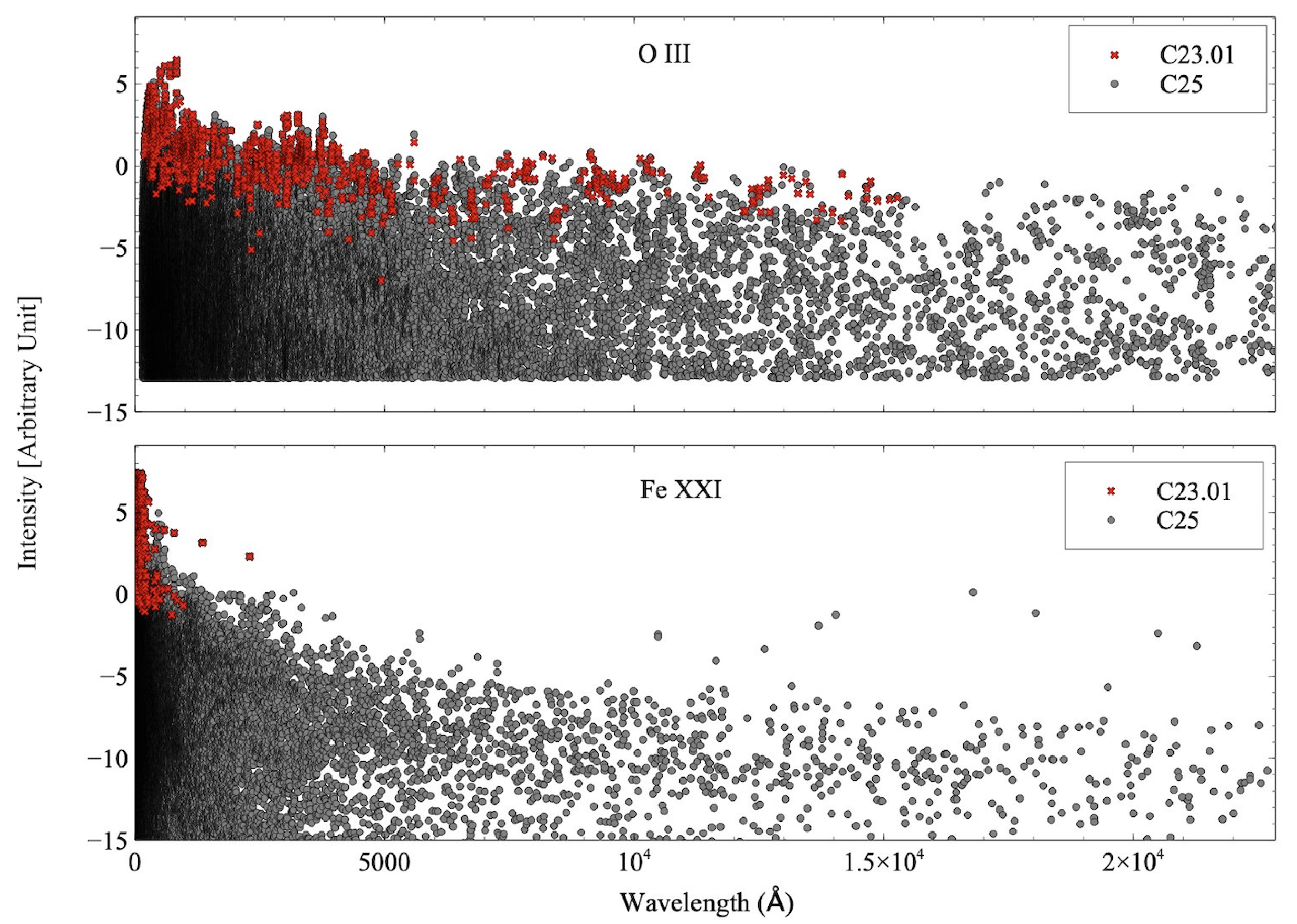}
      \caption{Comparison of predicted emission lines for O~III (top) and Fe~XXI (bottom) using the previous (C23.01, red) and updated (C25, gray) atomic datasets in \Cloudy{} (taken from Dehghanian et al., in preparation). The expanded and more complete atomic model used in the updated dataset results in a denser and more accurate distribution of predicted lines.}
         \label{fig:O3}
   \end{figure*}
\subsection{Chianti Atomic Database} 
The Chianti atomic database used by \Cloudy{} has been updated to the newer version, Chianti v10.1 \citep{1997A&AS..125..149D,2023ApJS..268...52D}. 
\citet{2022Astro...1..255G} introduced a script \url{https://gitlab.nublado.org/cloudy/arrack} that re-casts the latest Chianti database into the format used in Chianti v7 \citep{2012ApJ...744...99L}, which is the format compatible with \Cloudy{}. 
In the previous release of \Cloudy{}, we included a version of the Chianti database that contained only energy levels below the ionization potential, thereby excluding autoionizing levels. 
A complete version of the \Cloudy{} compatible Chianti v10.0 data \citep{2021ApJ...909...38D}, including these autoionizing levels, was available separately on \url{https://data.nublado.org/}. 
With the current update, \Cloudy{} now includes Chianti v10.1 with all 
levels fully 
integrated (including autoionizing levels). This expanded dataset is particularly important for applications in X-ray astronomy. 
The full database, as well as the version without autoionizing levels, of the \Cloudy{}-compatible Chianti v10.1 is available for download at \url{https://data.nublado.org/chianti/}.

\subsection{Updated H-like 2s Energies}
During the work described in \citet{2024arXiv241201606G} to resolve the $2p_{1/2,3/2} \rightarrow 1s_{1/2}$ doublet (hereafter Lyman $\alpha_{2,1}$), the Ly$\alpha_2$ line was found to overlap ambiguously with the $2s_{1/2} \rightarrow 1s_{1/2}$ M1 transition. This ambiguity arises because the upper levels of these transitions, $2p$ ($^2P_{1/2}$) and $2s$, are very closely spaced in energy. To disambiguate these lines, we updated the energy of the $2s$ levels of the H-like species to more accurate values obtained from \citet{2015JPCRD..44c3103Y}, thus clearly distinguishing M1 and the Ly$\alpha_2$ lines.

\subsection{Levels in Atomic Models}

Since H-like Fe Ly$\alpha$ lines are important X-ray diagnostics, we increase the number of resolved levels included in the H-like Fe\,{\sc xxvi} atom by default from 15 to 55 (i.e., all $nl$-resolved levels up to $n=10$). Increasing the number of levels enables a more refined computation of the collisional physics occurring in the higher energy levels. However, the size of the atomic models is restricted based on the computing time and available memory. 

Additionally, the number of collapsed levels of C\,{\sc vi}, N\,{\sc vii}, O\,{\sc viii}, Si\,{\sc xiv}, and Fe\,{\sc xxvi} have been increased from 5 to 15. The number of collapsed levels are relatively computationally inexpensive, since they are only $n$-resolved. So we have increased the number of collapsed for ions that contribute to large fraction of the gas physics. 

\subsection{Improvements to the output}

\subsubsection{Main output}

As a result of the work described in Section\,\ref{sec:npSplit}, for the lines whose energy separation is greater than a given spectral resolution (the default value for this resolution is 0.25~eV), \Cloudy{} now prints the $j$-resolved doublets, instead of the previous single line, under emergent and intrinsic line intensities.

\subsubsection{`save line labels' command}

The code has been changed to give the correct level indices in the \texttt{save line labels} output as they are defined in the input files for the database. In previous versions, it would give the level indices as they were stored in memory after reading the data files. In principle, this change can affect all databases, but in practice, this is only relevant for Chianti model atoms. This change will also affect the level indices used for line disambiguation.

Additionally, the ``extra'' Lyman lines now have two entries in the \texttt{save line labels} output one coming from the $j=1/2$ line stack and the other from $j=3/2$. The purpose of these lines is to fill in the ``gaps'' between the highest level and the continuum above, which arise due to the nature of having a finite H$^0$ model (also detailed in Section 3.1.4) of \citealp{2017RMxAA..53..385F}). As such, no changes are needed in their treatment.

\subsubsection{Line Labels}

With the need to distinguish, for example, M1 lines from Ly$\alpha$ lines, or various line components from the main emission line, there arose a need for more verbose line labels. So, the line labels in \Cloudy{} have been updated to use labels longer than our traditional use of four characters. However, these longer labels need to be in double quotes e.g. \texttt{"Fe 26 M1"}. Previously, we had line components that contributed to the lines with the following labels:
\begin{itemize}
\item 
\texttt{"Inwd"}---fraction of the line re-emitted toward the source,
\item
\texttt{"Pump"}---contribution to the total line intensity by continuum pumping,
\item
\texttt{"Coll"}---contribution to the total line intensity by collisional excitations i.e.\ the contribution to the gas cooling by this line,
\item
\texttt{"Heat"}---contribution to the gas heating by this line by collisional de-excitation of the upper level (this is a negative contribution to the line intensity).
\end{itemize}
All species used the same label. This made it difficult to identify what line that component contributed to. We have now disambiguated this by expanding the labels to now read e.g. \texttt{"H  1 M1 Pump"}, replacing the previous \texttt{"Pump M1"}.


\section{Molecular Data}
\label{sec:molecular_data}

\subsection{Temperature limits in UMIST chemistry}
\Cloudy{} primarily uses reaction rate coefficients from the UMIST Database for Astrochemistry (UDfA) \citep{2024A&A...682A.109M}. In UDfA, the rate coefficient for a two-body gas-phase reaction is given by a modified Arrhenius-type formula:
\begin {equation}
k=\alpha \left(\frac{T}{300}\right)^\beta \, \exp(-\gamma /T),
\end {equation}
where $T$ is the temperature of the gas. 
The UDfA provides fitted coefficients that are valid over specific temperature ranges. However, \Cloudy{} operates across a much broader range of temperatures--from the cosmic microwave background (CMB) up to 10$^{10}$~K, 
depending on the astrophysical environment. As a result, simply extrapolating these rate coefficients beyond their valid ranges can lead to unphysical values at both high and low temperatures. To prevent this, \Cloudy{} applies the following temperature caps:
Following \citet{2023RNAAS...7..153S}, for $\beta$ $>$  0, the rate is capped at high temperatures:
$k(T>$ 5000~K) = $k(T$ = 5000~K).
For $\beta$ $<$ 0, the rate is capped at low temperatures:
$k(T <$ 10~K) = $k(T$ = 10~K).
Similarly, for $\gamma$ $<$ 0, the rate is also capped at low temperatures to avoid divergence: $k(T <$ 10~K) = $k(T$ = 10~K) \citep{2011A&A...530A...9R}.
These caps ensure the stability and physical plausibility of reaction rates in the wide range of temperatures modeled in \Cloudy{}.

\subsection{Si-chemistry}
We have extended our existing silicon-chemistry network \citep{{2022ApJ...934...53S},{2023RNAAS...7...45S}}, which now includes 21 Si-bearing species: SiS, HSiS, HSiS$^+$, SiS$^+$, SiC, SiC$^+$, SiC$_2^+$, SiNC$^+$, SiH$_2$, SiCH$_2$, SiCH$_2^+$, SiNC, SiN, SiN$^+$, SiO$^+$, SiC$_2$, SiH$_2^+$, SiH, SiOH$^+$, SiO, and SiO$^+$. Notably, the reaction N + SiC$^+$ $\rightarrow$ Si$^+$ + CN significantly impacts the column density and line intensity of CN. Among these 21 Si-bearing molecules, line intensities have been predicted for SiS, SiO, and SiC$_2$. The corresponding energy levels and collisional rate coefficients for these molecules are adopted from the LAMDA Database \citep{2005A&A...432..369S}. 

\subsection{Ti-chemistry and TiO}
TiO is the dominant source of opacity in the atmosphere of cool stars \citep{Lodders_2002}, and it is observed in the stellar atmosphere of M-type giant stars \citep{2013A&A...551A.113K} as well.
However, it is not observed in the ISM due to the high depletion of Ti. In environments where dust is not present, TiO will be observed in the gas phase.
Recently, we have added 229 Ti-related reactions in the chemical network \citep{2024RMxAA..60..373S}. However, there is a scarcity of reaction rates for Ti-chemistry. 
Hence, we have incorporated some reactions
that are available in UDfA, \citet{2021ApJ...923..264T}, \citet{1980AJ.....85.1382C}. 
For reactions not directly available, we scaled analogous silicon-based reactions from UDfA.  
In addition, we consider 230 fine-structure energy levels, the corresponding 223
radiative transitions, and 444
collisional transitions with ortho and para H$_2$
and predict 66 TiO lines. Further details are available in \citet{2024RMxAA..60..373S}.

We modeled the circumstellar envelope of the oxygen-rich red supergiant VY Canis Majoris to validate our update. Our model reproduces the observed column density of TiO. We notice that, in the gas-phase, Ti is mainly in TiO for temperatures above 1400 K, and TiO$_2$ dominates at a lower temperature. 

Note that 
Ti-chemistry is not enabled by default. 
Tests show that our linear algebra package
can have convergence
problems under some extreme conditions when
TiO is included.
To activate the chemistry , use the command \cdCommand{set chemistry TiO on}.

\subsection{Phosphorous bearing molecules}
Phosphorus (P) is essential for the formation of complex compounds, including DNA and RNA, which are fundamental to life. P-bearing molecules have been observed in the Milky Way \citep{2022FrASS...9.9288R,2020MNRAS.492.1180R, 2019MNRAS.489.4530F, 2020A&A...633A..54C}, as well as in extra-Galactic environments \citep{2022A&A...659A.158H}. We have updated the gas-phase chemical reaction rates and molecular
lines for P-bearing molecules in the spectral synthesis code \Cloudy{}. 
The corresponding molecular reaction rates were obtained from UDfA.
As a result, we predict column densities of 14 P-bearing molecules, PH, PH$^+$, PH$_2$, PH$_2^+$, PH$_3$, PH$_3^+$, CP, CP$^+$, HCP, HCP$^+$, PN, PN$^+$, PO, PO$^+$.
Among these, we predict molecular lines for PN, PO$^+$, PH$_3$. The energy levels and radiative and collisional rates for PO, PN, PO$^+$, and PH$_3$ from the LAMDA database. 

\begin{figure*}[t]
    \centering
    \includegraphics[width=0.95\textwidth]{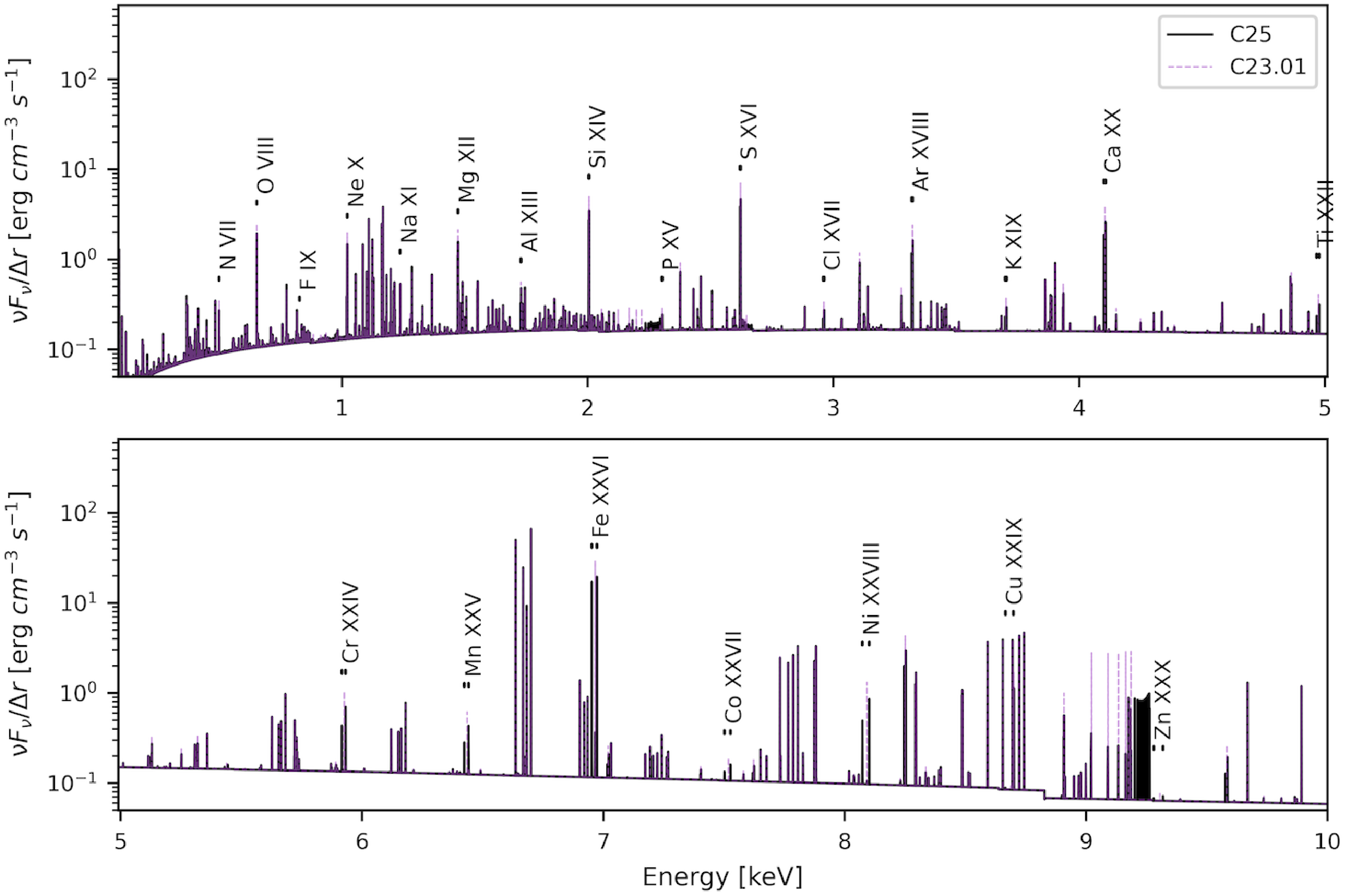}
    \caption{Taken from \citet{2024arXiv241201606G}, \Cloudy{} simulated spectrum of the Perseus cluster core, showing the $j$--resolved Lyman $\alpha$ lines in C25. Only the energy range, $0.4$ - $10$ keV, is covered by the XRISM mission.}
    \label{fig:npSplit_SpectralPlot}
\end{figure*}
\section{\Cloudy{} for microcalorimeter X-ray Astronomy}
\label{sec:xray}

\subsection{One-electron Lyman Doublets}
\label{sec:npSplit}

With the advent of X-ray microcalorimeters with spectral resolution $R\equiv E/ \Delta E>1200$ at 5.9~keV, like the one on XRISM, X-ray observations can now resolve the fine-structure doublets of Ly$\alpha$ lines of one-electron species for the first time in astrophysical plasmas (for the Sun this doublet was already resolved prior to XRISM). Although \Cloudy{} was not originally designed for high-resolution X-ray spectroscopy, the work in \citet{2024arXiv241201606G} has expanded \Cloudy{}'s treatment of one-electron systems to resolve the H-like Ly$\alpha$ doublets. Earlier work expanded on the two-electron system \citep{2020ApJ...901...68C, 2020ApJ...901...69C, 2021ApJ...912...26C, 2022ApJ...935...70C}, included in the C23 release \citep{2023RMxAA..59..327C}.

Figure~\ref{fig:npSplit_SpectralPlot}, taken from \citet{2024arXiv241201606G}, presents a model of the core of the Perseus cluster. 
This model is a collisionally ionized plasma at 
$T_e = 4.7\times10^7$ K, and 
$n_\mathrm{H}=10^{-1.5}$ cm$^{-3}$.
The goal here was to resolve the single $np\rightarrow1s$ lines predicted by \Cloudy{}, into their fine-structure j-resolved doublets, $np_{1/2}\rightarrow1s_{1/2}$ and $np_{3/2}\rightarrow1s_{1/2}$. This increases \Cloudy{}'s X-ray spectral resolution to match that of XRISM.
Part of the challenge was to retrofit the H-like fine-structure doublets into \Cloudy{}'s already existing full collisional radiative model (hereafter CRM) solver. Previously, \Cloudy{} made use of psuedo-states to represent closely-space Rydberg levels at high principal quantum numbers \citep{2017RMxAA..53..385F}, to reproduce the classical case B intensities of H and He recombinations lines. The psuedo-states were replaced with models of higher-$n$ shells, as computers became faster. \Cloudy{} now employs $nl$--resolved states for low $n$, and ``Collapsed states'' that are not $l$--resolved for high $n$.

The ``extra'' Lyman line arrays in \Cloudy{} have been expanded to include the treatment of j-resolved Lyman lines of one-electron species, in addition to their original purpose. 
We left the framework of the He-like ``extra'' Lyman lines unchanged. 
We begin by calculating the one-electron $np$ energies as described in  \citet{2024arXiv241201606G}. 
We then approximate the $j$-resolved population densities of these lines to the ratio of their statistical weights,
\begin{equation}
    n_{npj} = 
    \begin{cases}
    n_{n} \left(\frac{g_{np}}{2n^2} \frac{g_{npj}}{g_{np1/2}+g_{np3/2}}\right), & {\rm collapsed\ states}\\
    n_{np} \left(\frac{g_{npj}}{g_{np1/2}+g_{np3/2}}\right), & {\rm resolved\ states}
    \end{cases}
\end{equation}
where $g_i$ is the statistical weight of level $i$. 
Ideally, we would be able to derive their population densities using the CRM solver. The lack of reliable proton-impact $j$-changing collisional data for most one-electron species, makes the aforementioned approximation the best presently available solution. 

\begin{figure}[t]
    \centering
    \includegraphics[height=1.3\columnwidth]{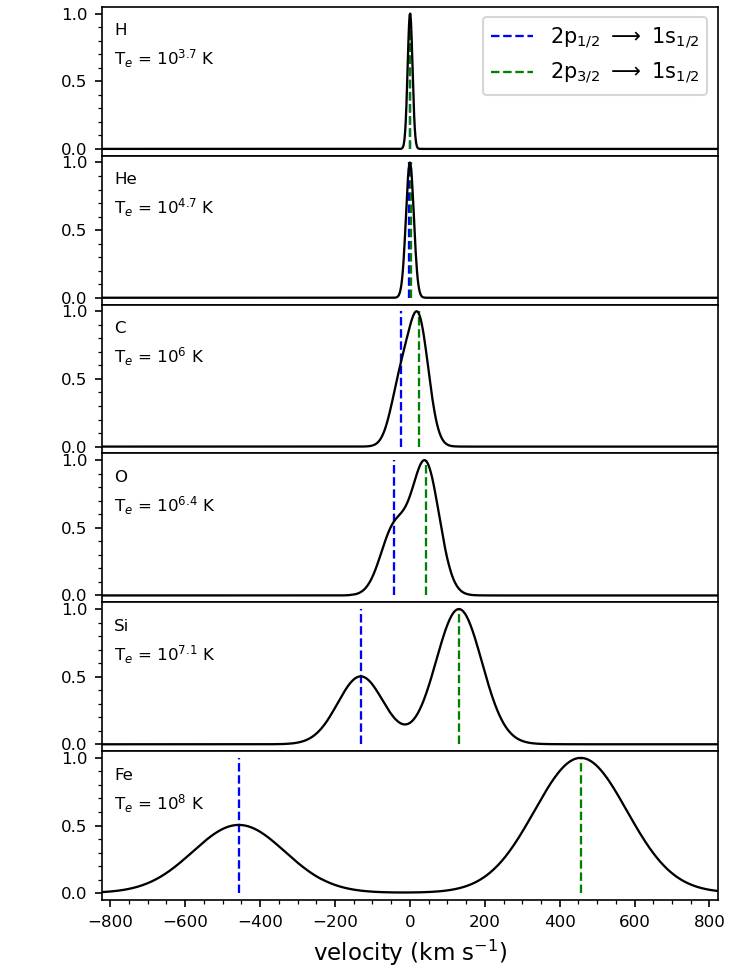}
    \caption{Taken from \citet{2024arXiv241201606G}, this plot shows normalized line opacities against line-of-sight velocity for various one-electron 2$p$ fine-structure doublets at a spectral resolution of 0.25 eV. The blue and green dashed lines mark the positions of the Ly$\alpha_{j=1/2}$ and Ly$\alpha_{j=3/2}$ components, respectively. Each panel's top left corner indicates the gas temperature, chosen to match the peak abundance of the ion under conditions of collisional equilibrium. As nuclear mass increases, the line profiles generally become narrower, while higher temperatures cause them to broaden.}
    \label{fig:LineOverlap}
\end{figure}
At the default \Cloudy{} spectral resolution (further details in \ref{sec:database_lyman_extra_res}), a few low-$Z$ ($Z<8$) Ly$\alpha_{1}$ and  Ly$\alpha_{2}$ lines overlap. Figure~\ref{fig:LineOverlap} shows line opacities as a function of line-of-sight velocity for the $j$--resolved Lyman $\alpha$ doublets. Evaluated at temperatures where that ion is most abundant in collisional ionization equilibrium, the figure illustrates how the doublet splitting grows with increasing atomic number ($Z$). 
By default, we report the total line intensity for first- and second-row elements, as well as the individual members of the multiple for heavier elements.
It is only for the third row and heavier elements that the 
lines become two non-overlapping lines at the default spectral resolution. For H-like Ly$\alpha$ lines the code uses the theory in \citet{1980ApJ...236..609H} to calculate the escape probability $\beta$, using the line opacities from the coarse continuum. \Cloudy{} computes nebular spectra using multi-grid methods with two continua: a coarse continuum for overall radiation and continuous processes, and a fine continuum for detailed line transfer and line overlapping \citep[further detail on the fine and coarse continua are given in][]{2005ApJ...624..794S}. This theory implicitly assumes a single line with a Voigt profile. We update \Cloudy{}'s escape probability to use the fine opacity mesh instead, which allows for the treatment of overlapping lines (further detailed in \citealp{2024arXiv241201606G}). As a result of this update, we no longer allow users to disable the calculation of the fine mesh, and the \texttt{no fine opacities} command has been removed.

\citet{2025A&A...694L..13G} found that the work described above yielded a novel result: at hydrogen column densities, $N$(H), above ~$10^{22}$ cm$^{-2}$ of the X-ray emitting gas, the ratio of the Ly$\alpha_1$ to Ly$\alpha_2$ ratio deviates from the expected 2:1 ratio in the optically-thin limit \citep{1986PASJ...38..225T}. Larger column densities correspond to larger optical depths. So this deviation arises from changes in the optical depths of the j-resolved components of Ly$\alpha$, which reflect the hydrogen column density of the associated gas. Further details on this physics, the above work and a novel $N$(H) diagnostics are described in \citet{2025A&A...694L..13G} and \citet{2024arXiv241201606G}.

\subsubsection{H-like Lyman Resolution Command}
\label{sec:database_lyman_extra_res}

Well-known quantum mechanical theory gives us that the fine-structure splitting i.e. the difference between the energy levels ${np(^2P_{1/2})}$ and ${np(^2P_{3/2})}$, is of the order $Z^2(n-1)/n^2$ \citep{1957qmot.book.....B}. So we need increasing resolving power to resolve the fine-structure lines with increasing $n$ and decreasing $Z$. The current (XRISM) and future (NewAthena) planned microcalorimeter observatories have spectral resolution $R$ of 5 eV and 2.5 eV, respectively. So by default we set \Cloudy{}'s one-electron fine-structure line spectral resolution to $2.5$ eV $/10=0.25$ eV. Consequently, we introduce a new command allowing the user to alter this default resolution,\\
\texttt{Database H-like Lyman extra resolution} \textit{R},
\\
where, \textit{R} $=$ 0.25 eV by default.
Note, that we do not allow H and He Lyman lines to be resolved into their fine-structure doublets regardless of the user set resolution for two important reasons: a) this will break important physics needed by the \Cloudy{} solvers, b) current and future known instruments will not be able to resolve these lines.

\subsection{Physics of the 1 keV blend}

\begin{figure*}
\centering

\begin{minipage}{0.8\textwidth}
    \centering
    \includegraphics[width=\textwidth]{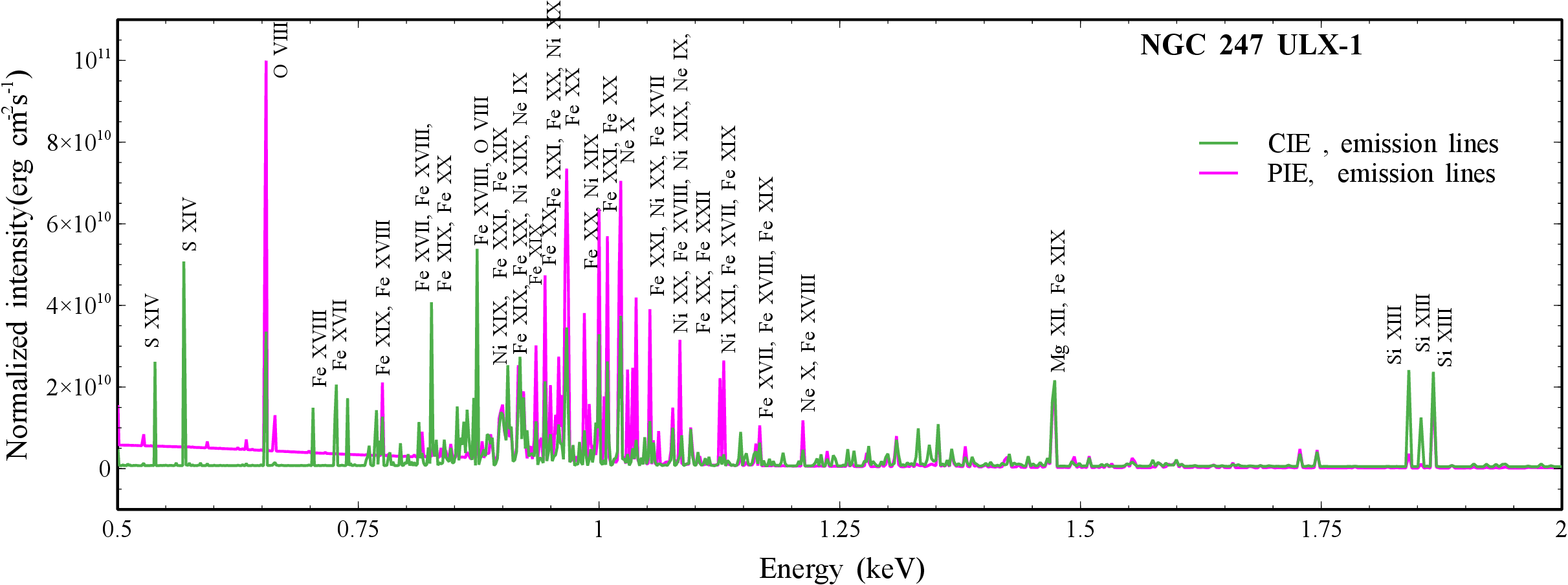}
\end{minipage}


\begin{minipage}{0.8\textwidth}
    \centering
    \includegraphics[width=\textwidth]{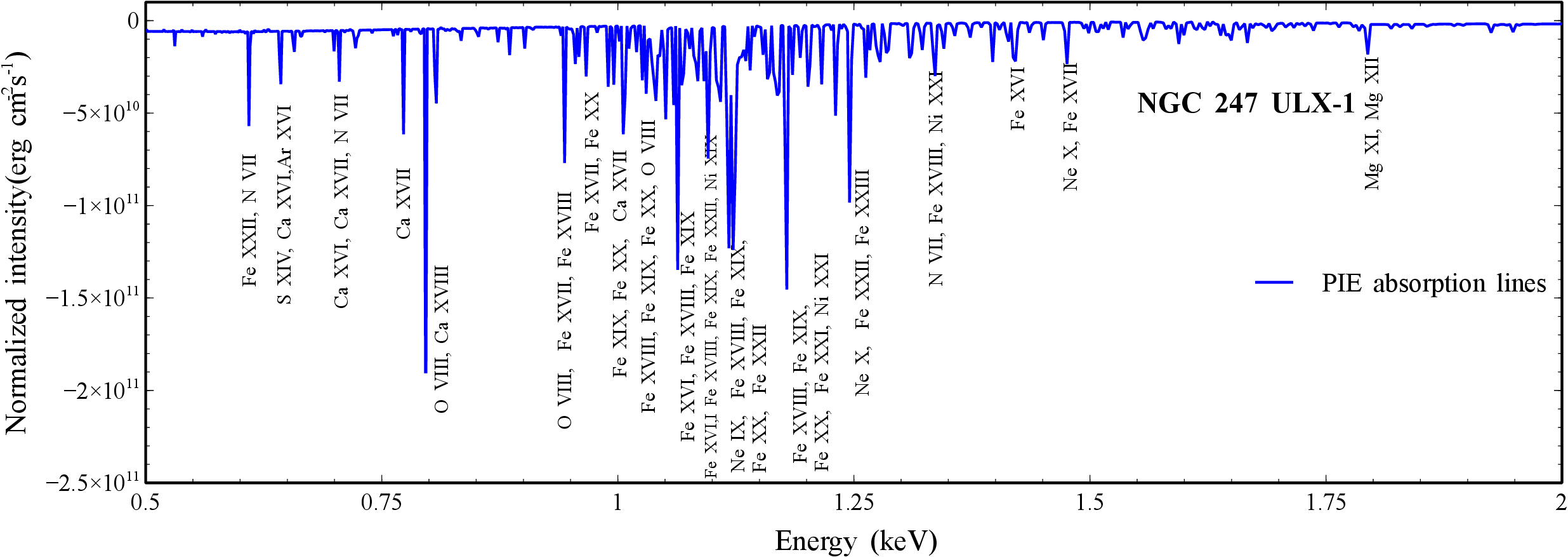}
\end{minipage}


\begin{minipage}{0.4\textwidth} 
    \centering
    \includegraphics[width=\textwidth]{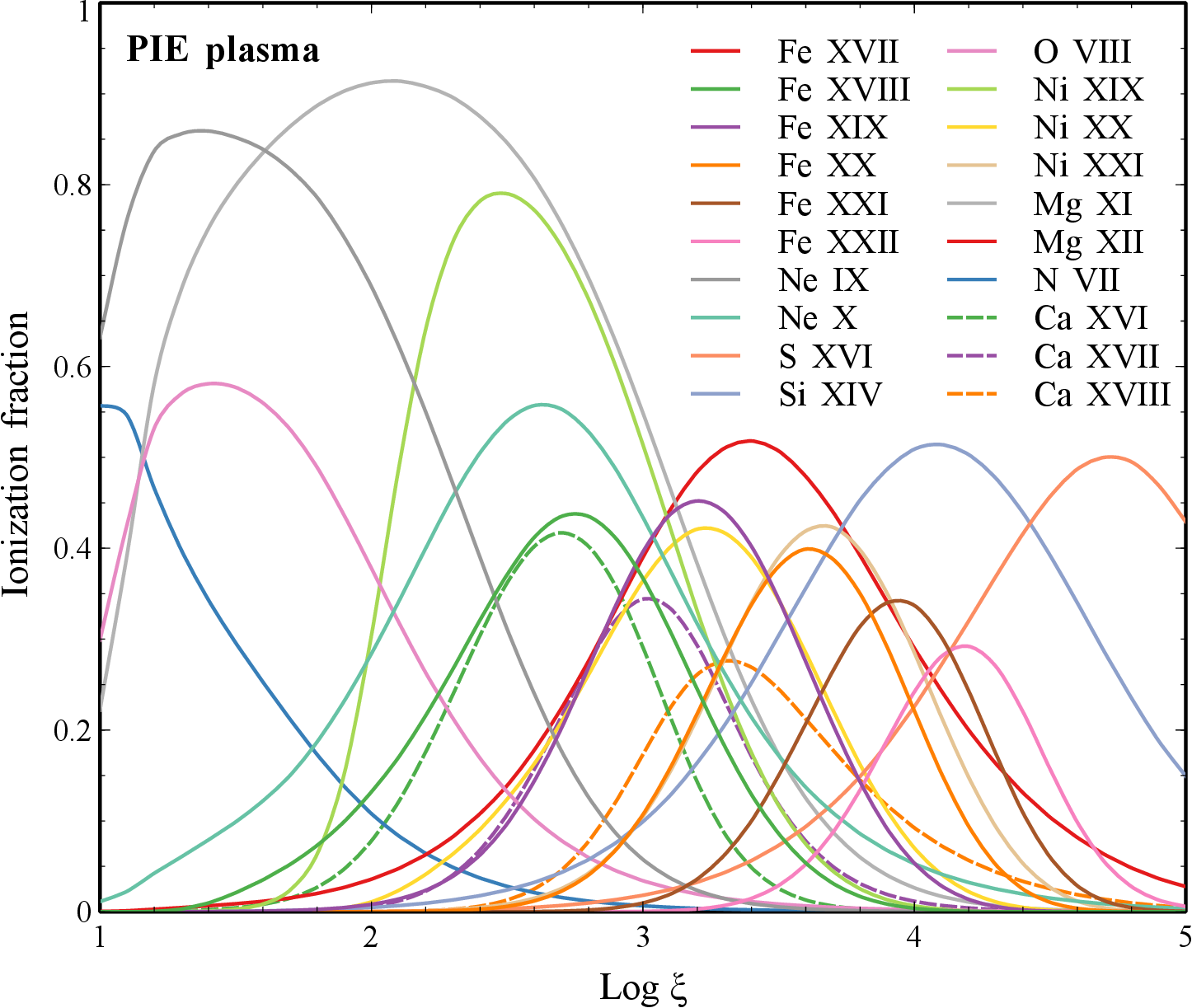}
\end{minipage}
\begin{minipage}{0.4\textwidth}
    \centering
    \includegraphics[width=\textwidth]{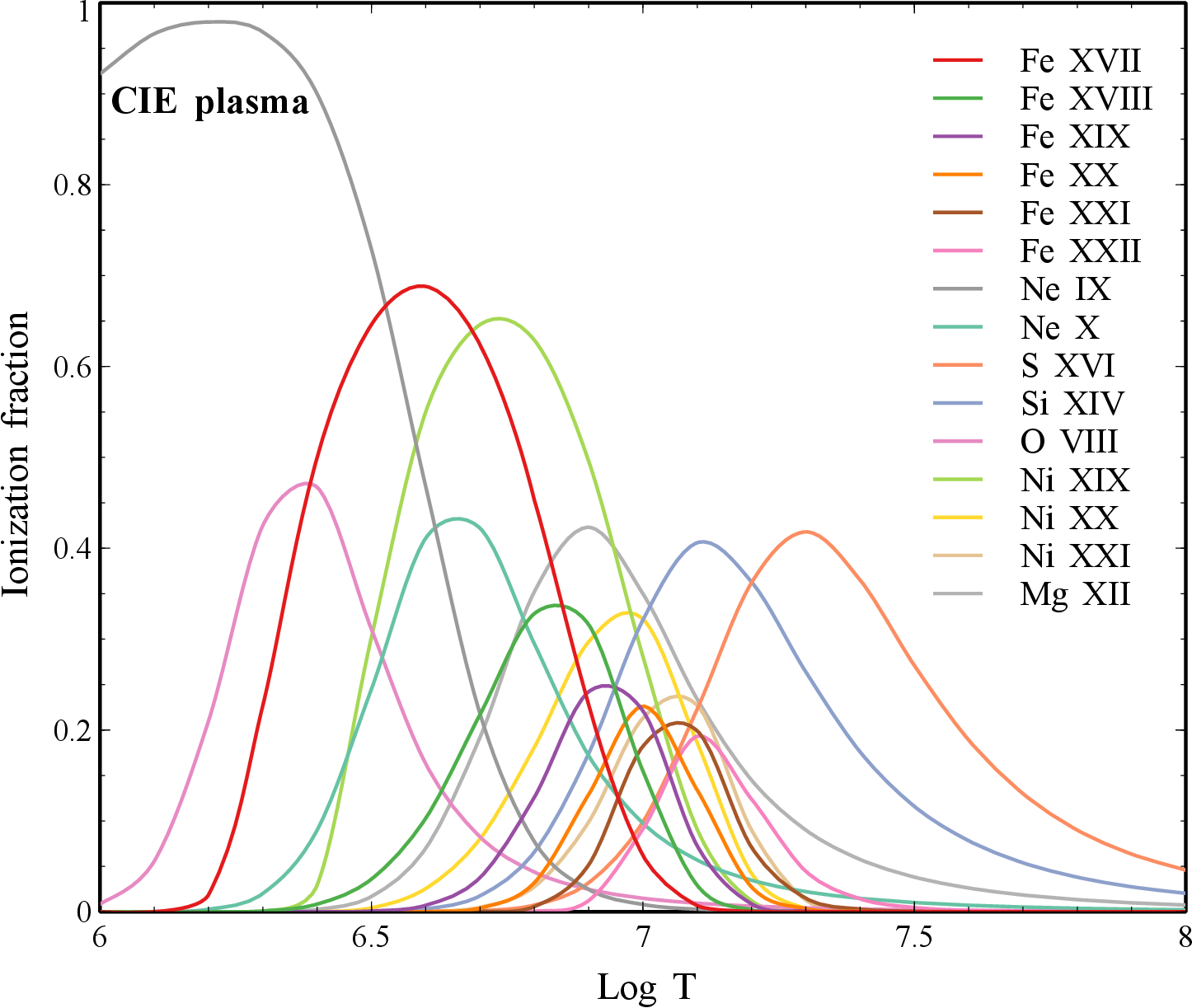}
\end{minipage}

 \caption{Taken from \citet{2024arXiv240702360C}. High-resolution \Cloudy{} model of the 1 keV feature for NGC 247 ULX-1 at the spectral resolution of \textit{NewAthena}. 
Top panel: Photoionized (PIE) and collisionally-ionized (CIE) emission lines at the best-fit values reported in \citet{2024arXiv240702360C}. 
Middle panel: PIE absorption lines based on the same best-fit model.
Bottom left: Charge state distribution for the PIE plasma. 
Bottom right: Charge state distribution for the CIE plasma.}

\label{fig:hires}

\end{figure*}

\subsubsection{Introducing ``mixed'' command}

Until the previous version, \Cloudy{} used experimental energy values from the Chianti database by default, due to their superior accuracy \citep{2013MNRAS.429.3133L}. In the current version, theoretical energy values can now be incorporated in cases where experimental data are absent.
The \Cloudy{} command to use such a 
 ``mixed'' case is: \texttt{database Chianti mixed},
which was introduced in \citet{2024arXiv240702360C}
and used to explain the origin of the 1 kev feature in X-ray binaries (XRBs).\\

\subsubsection{1 keV Line Blend}

In several instances, spectrometers measure the integrated flux over a defined energy range, which often prevents the unambiguous identification of individual line contributions within blended spectral features. The \texttt{Blnd} command in \Cloudy{}, first introduced in \citet{2017RMxAA..53..385F}, reports the total flux from the 1 keV line blend, matching what is observed.

A well-known case is the ``1 keV feature'' in XRBs, where residuals frequently appear between 0.5 and 2 keV due to unresolved line blends which varies both in centroid and intensity across various types of X-ray binaries as well as over time within the same binary \citep{2002ApJ...579..411P, 2006MNRAS.368..397S, 2015MNRAS.454.3134M, 2020MNRAS.494.6012W}. Despite numerous modeling efforts using a range of physical scenarios, a comprehensive scientific explanation for the origin and variability of the 1 keV feature remained elusive.  

\citet{2024arXiv240702360C} used the \texttt{set blend} command to construct a line blend using all the lines within the energy range 0.5-2.0 keV.
This blend was introduced in \texttt{blends.ini} with the name \texttt{Blnd 11}.
The sensitivity of the flux of this line blend was tested against the spectral energy distribution (SED) shape, ionization parameter ($\xi$), column density ($N_{\mathrm{H}}$), and gas temperature ($T$), following the methodology described in \citet{2020ApJ...901...68C, 2020ApJ...901...69C, 2021ApJ...912...26C, 2022ApJ...935...70C} to probe the physical origin and  spectral variability of the 1 keV feature.

\subsubsection{\textit{NewAthena} predictions}

Using the 1 keV blend, \citet{2024arXiv240702360C} conducted a thorough analysis of emission
and absorption lines under three specific conditions: photoionization
equilibrium (PIE), collisional ionization equilibrium (CIE), and reflection of X-rays off the inner regions of an accretion disk.
The 1 keV blend was systematically varied with respect to ionization parameter, temperature, column density, and the shape of the SED for five XRBs: two ultraluminous X-ray sources (ULXs), NGC 247 ULX-1 and NGC 1313 X-1; one X-ray pulsar, Hercules X-1; and two low-mass X-ray binaries (LMXBs), Cygnus X-2 and Serpens X-1. The \textit{XMM-Newton/RGS} and \textit{NICER} spectra of these sources were fit using \Cloudy{} models incorporating the newly implemented 1 keV blend.
A self-consistent framework was established to explain the variability of the 1 keV spectral feature, with its diversity attributed to variations in SED shape, ionization state, temperature, column density, and disk reflection properties.

Figure \ref{fig:hires} shows a \Cloudy{} model of the 1 keV blend in NGC247 ULX-1, based on the SED and best-fit parameters from \citet{2024arXiv240702360C} at the spectral resolution of \textit{NewAthena}. This model quantifies the atomic line contributions to the spectrum, including the newly implemented 1~keV line blend in \Cloudy{}. The spectrum
has been decomposed into its individual CIE and PIE components, with
strong lines from  within the 1 keV blend labeled for clarity.

\subsection{Updated Inner Shell Energies}

We updated the K$\alpha$ and K$\beta$ fluorescence lines energies from the original \citet{1993A&AS...97..443K} data based on the corrections and prescriptions described in the SRON-SPEX ``Atlas of Innershell Ionization lines''\footnote{https://var.sron.nl/SPEX-doc/physics/trpb04c.pdf}, which relies on the more accurate data from \citet{1969ApJS...18...21H} and \citet{1967RvMP...39..125B}. In particular, we verified that the K-shell transition energies for Fe~II to Fe~XXII are now in very good agreement with the more recent calculations by \citet{2003A&A...403.1175P, 2004A&A...414..377M, 2004A&A...418.1171B}. For S and Si, the values remain based on experimental data, as introduced in the patch by \citet{2021RNAAS...5..149C}.

\subsection{XRISM/\textit{Resolve}--specific Initialization File}

X-ray microcalorimeters pose unique challenges and opportunities. With the launch of XRISM \citep{2025PASJ..tmp...28T}, these spectra have become a reality.
The \Cloudy{} team participated in a XRISM-focused \Cloudy{} workshop in Tokyo in the Summer of 2024. A pre-release version of the code was exercised by several dozen JAXA scientists and students. The team worked to improve the code and interesting results came out \citep{2025A&A...694L..13G, 2025PASJ..tmp...30T}.

The spectral needs of an X-ray microcalorimeter are unique. 
We added an instrumentation-specific initialization file, \texttt{XRISM.ini} to the distribution data directory.
It increases the continuum spectral resolution and increases the number of levels included in models of 11-- through 1--electron iron. The predictions for a simulation of the Perseus cluster are shown in Figure \ref{fig:XRISMini}.

The higher-fidelity simulation took 50\% longer than the simulation with our default state. Its higher spectral resolution is obvious, as is the far larger number of lines.
The insights resulting from the microcalorimeter revolution is obvious.

 \begin{figure}
   \centering
   \includegraphics[width=\columnwidth]{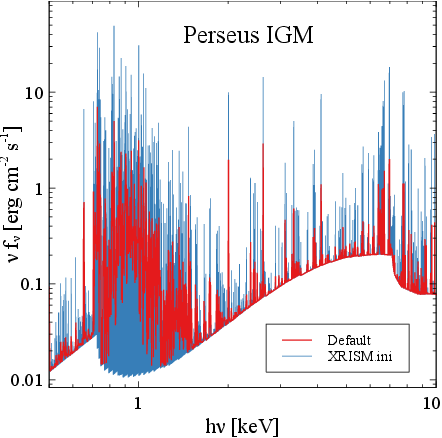}
      \caption{This compares our predicted spectrum of the Perses Cluster cooling flow as modeled by
      \citet{2020ApJ...901...68C}. It is the same simulation but compares our default setup with the use of our \texttt{XRISM.ini} 
      initialization file. }
         \label{fig:XRISMini}
   \end{figure}

\subsection{Updated Fe K Blends}

Up until the C23.01 release, \Cloudy{} had included Fe K lines heavily utilized by the X-ray community. With this release, the \texttt{FeK1} an \texttt{FeK2} lines, which were the one-electron and two-electron K$\alpha$ lines, have been replaced by \texttt{"Blnd" 1.77982} and \texttt{"Fe25" 1.85040A} respectively. The former is now defined as a blend of the following three lines:\\
\texttt{"Fe26" 1.77802A}\\
\texttt{"Fe26 M1" 1.78330A}\\
\texttt{"Fe26" 1.78344A}\\
Additionally, we have removed the following Fe K lines: \texttt{"FeKH"} which were fluorescent hot iron lines from Fe\,{\sc xviii}-Fe\,{\sc xxiii}, \texttt{"FeKC"} which were fluorescent hot iron lines from Fe\,{\sc xvii} as these are relics from early X-ray astronomy and are no longer relevant in modern studies. Lastly, no changes were made to \texttt{"FeKG"}, the grain production component of cold Fe.

\section{Miscellaneous improvements}
\label{sec:misc_impovements}

\subsection{H Ly$\alpha$ escape and destruction probability}

In the C23.01 release, we revised our calculation of the H Ly$\alpha$ destruction probability. For details on the updated escape and destruction probability treatment --- denoted as $\beta_{\rm HK}$ --- see \citet{2023RNAAS...7..246G}.
As  this sub-release was not accompanied by a full review paper, we take this opportunity to outline the resulting changes to the H Ly$\alpha$ physics, which remain relevant in the current release, C25.00.

\begin{figure*}
    \centering
    \includegraphics[width=\textwidth]{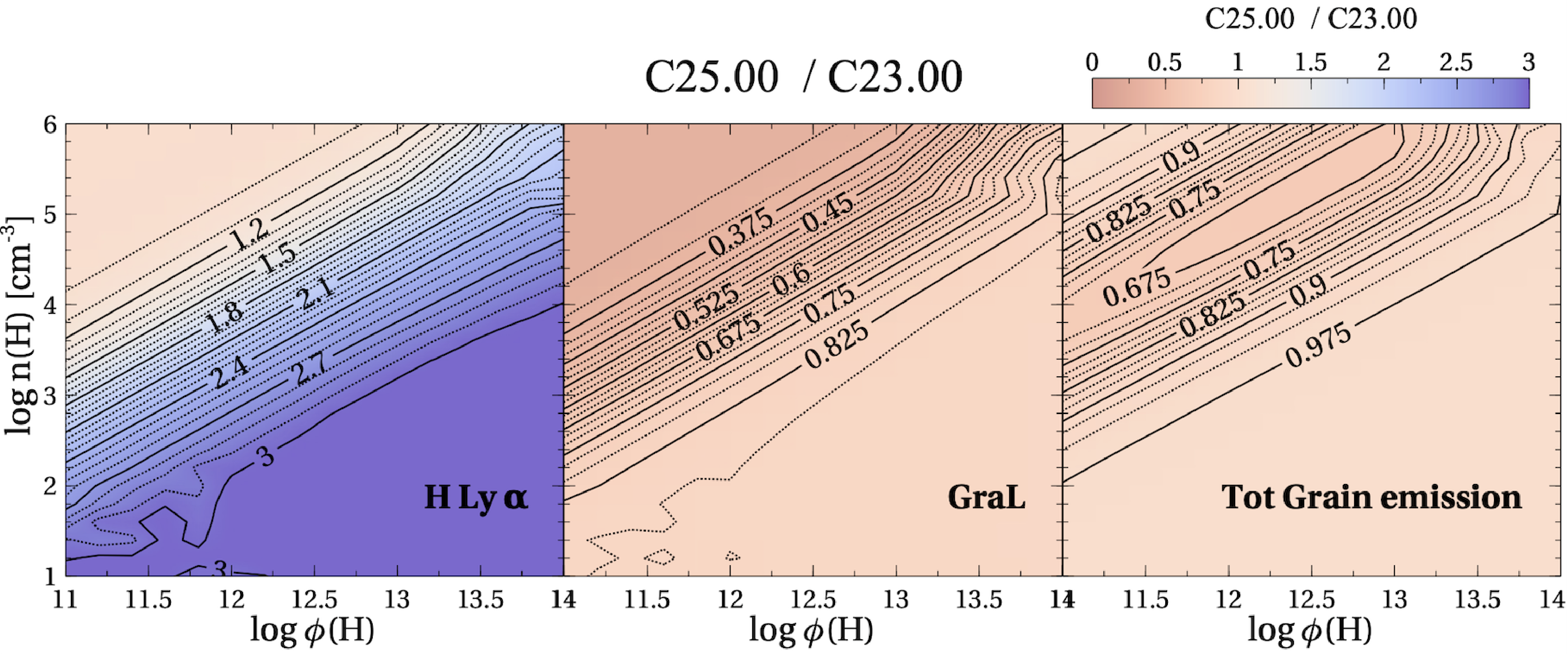}
    \caption{A contour plot of physical parameters predicted by C23.01 relative to the same quantity from C23.00, 
    for the baseline model \texttt{orion\_hii\_open.in} in the \Cloudy{} test suite. 
    The panels give, 
    \textit{\textbf{Left:}} taken from \citet{2023RNAAS...7..246G} shows H Ly$\alpha$, 
    \textit{\textbf{Middle:}} grain heating due to Ly$\alpha$  destruction, 
    \textit{\textbf{Right:}} total grain heating by all sources, lines, collisions, and incident continuum. 
    The ratio of ionizing photon flux $\phi(H)$ to hydrogen density $n(H)$ is effectively the ionization parameter $U$. The lower-right corner of the panels corresponds to high $U$, and the upper-left corner is low $U$. }
    \label{fig:contour}
\end{figure*}

This updated treatment led to noticeable changes in grain emission, particularly that associated with H Ly$\alpha$ absorption by grains. Figure~\ref{fig:contour} presents contour plots of the ratio of physical parameters predicted by C23.01 to those from C23.00, using a benchmark H\,{\sc ii} region model. The three panels show the H Ly$\alpha$ line intensity (left), grain heating from Ly$\alpha$ photon destruction (middle; hereafter \texttt{"GraL"}), and total grain heating from all sources
(right).

Dust grains are the main opacity source that absorbs and destroys Ly$\alpha$ photons in H\,{\sc ii} regions \citep{1978ppim.book.....S}. With the updated $\beta_{\rm HK}$, more Ly$\alpha$ photons now escape,
reducing the fraction absorbed by dust and thereby lowering grain heating from Ly$\alpha$ destruction (middle panel). Grains are heated by three main mechanisms: (i) the incident radiation field, (ii) collisions with gas particles, and (iii) absorption of line photons such as Ly$\alpha$. The total grain emission (right panel), reflects the combined effects of all three, and is reduced slightly due to the decreased Ly$\alpha$ absorption.

However, the total grain emission changes less dramatically than the grain heating by H Ly$\alpha$ because other processes, especially the incident starlight, also contribute. Thus the effects on grain emission is more subtle. Even in H\,{\sc ii} regions with moderate dust optical depths, dust can significantly hinder the escape of Ly$\alpha$ photons \citep{2011piim.book.....D}. 

The impact of the reduced Ly$\alpha$ destruction is most pronounced at low ionization parameters ($U$), defined as the ratio of the ionizing photon flux $\phi(H)$ to the hydrogen density $n(H)$. So, in the figure, high $U$ corresponds to the lower-right corner of the panels, and low $U$ to the upper-left. \citet{1998PASP..110.1040B} demonstrated that grain absorption depends on $U$. 
At low $U$, enhanced opacity in the 1s$\rightarrow$2p transition increases the probability that a Ly$\alpha$ photon is absorbed near its point of origin (a rate referred to as ``on-the-spot''; hereafter OTS rate). Since GraL is directly proportional to the OTS rate, the revised destruction probability leads to a significant reduction in both the OTS rate and GraL, particularly at low $U$.

We find a local minimum of the total grain emission that occurs at $\frac{\phi(H)}{n(H)}\sim 10^{7.2}$ photons\,cm\,s$^{-1}$ (right panel). This arises largely due to the fact that both diffuse field heating and gas-grain collisional heating reach local minima in their relative deviations under the new $\beta_{\rm HK}$ prescription. These ``dips'' compound the overall reduction in grain heating efficiency in this low $U$ regime.

Despite these changes to Ly$\alpha$ and grain physics, the new $\beta_{\rm HK}$ has little effect to most of other observable emission lines. In particular, the classical BPT spectral lines, [O\,{\sc iii}]$\lambda$5007, [N\,{\sc ii}]$\lambda$6583, [S\,{\sc ii}]$\lambda\lambda$6716,6731, and [O\,{\sc i}]$\lambda$6300 are minimally affected.


\subsection{Numerical Methods}

A version of the GTH Algorithm \citep{GTH85,Zhao2020}, which guarantees the
positivity of equilibrium solutions of Markov chains by a specific ordering
of operations in Gaussian elimination, has been applied to the atomic level
solver in cases where there are no outside sources from other ionization
states or chemical processes.
This has addressed an infrequent, but longstanding, code failure mode where
negative level abundances were predicted in species such as Ca\,{\sc i}, and has
in general been found to give more accurate results for levels with trace
populations.

The dynamical solver has been updated to gracefully
handle temperature floors reached in cooling calculations of
a recombining gas.
Temperature floors may be reached by a gas exposed
to intense photoionization,
e.g., in the vicinity of a quasar \citep{2025Natur.638..360R},
or by an extremely rarefied gas exposed to
intense cosmic radiation.
In both cases, the externally deposited energy forces 
the gas to come to equilibrium at a certain temperature,
and may prevent it from reaching the temperature prescribed with
the \texttt{stop time when temperature below} command.
Previously, the solver would continue integrating
the evolution of the system to unphysical time-spans,
or it could even crash.

\subsection{Physical constants}

The physical constant have been updated to the Codata 2022 values \citep{2024arXiv240903787M}.

\section{Infrastructure Changes}
\label{sec:infra}

\subsection{The C++ standard}

The language standard used by \Cloudy{} has been changed from C++11 to C++17. This changes the minimum requirements for the compilers that we support. For GNU/g++ you now need version 8 or later, while for LLVM/clang++ you need version 7 or later.

The Oracle Studio compiler has not been maintained for a long time and does not support C++17. Support for this compiler has been dropped.

\subsection{API changes}

The API for the routines {\tt cdLine()} and {\tt cdEmis()} has been changed. This may affect programs calling \Cloudy\ as a subroutine. When using a wavelength parameter in these calls, it now has to have type {\tt t\_wavl}. This allows the user to indicate whether the wavelength is in air or vacuum.

\subsection{Parser changes}

The parser now restricts the use of non-ASCII characters in scripts. They are forbidden in the command part, but are still allowed in comments. The code now aborts if they are found where they don't belong. Before this change, the parser would simply skip non-ASCII characters, which can lead to obscure errors. One example is when the number $-4$ is typed with the unicode math minus symbol. The unicode minus symbol would be skipped and the number would be read as 4 rather than $-4$.

\subsection{Changed commands or options}

\Cloudy{}~C22 introduced line disambiguation \citep{2023RMxAA..59..327C}. At that time two commands were overlooked: the \texttt{normalize} and \texttt{stop line} commands. Support for line disambiguation has now been added for these two commands.

The \texttt{print line vacuum} command has been fixed. In the wake of that fix, several changes have been implemented that allow better handling of air and vacuum wavelengths. First of all, the code now always uses vacuum wavelengths internally and only converts to air wavelengths right before the numbers are printed (that was not the case in older versions of \Cloudy{}). The conversion will only be done for spectroscopic lines and not for continuum wavelengths (e.g., the \texttt{save continuum} output will always use vacuum wavelengths if wavelength units are requested - this behavior is {\em not} new). Several commands have now been amended to allow optional keywords \texttt{air} or \texttt{vacuum} to indicate the type of the wavelength. This forces the interpretation of the number, regardless of whether the \texttt{print line vacuum} command was used or not. Everywhere line disambiguation is supported, these new keywords are also supported. Additionally the following commands now accept these keywords: \texttt{set blend} (for the wavelength of the blend itself as well as the components of the blend), \texttt{print line sort wavelength range}, and \texttt{monitor Case B range} (for the wavelength range).

The \texttt{print path} command has been improved. It now accepts an optional string between double quotes that will be used as a wildcard pattern to match specific data files. Note that the standard C++ ECMAScript grammar for regular expressions will be used, {\em not} the familiar wildcard characters that most UNIX shells use. This command now immediately exits, making it more convenient to find data files.

The \texttt{table star available} command will now detect all SED grids, including user-defined grids. The output of this command has been completely redesigned. The \texttt{compile stars} command (without additional options) will now also work on user-defined grids.

The keyword \texttt{quiet} for the \texttt{set blend} command has been improved and will quietly ignore the blend if any of the blend components cannot be found.

The \texttt{illumination} command has not changed, but its description in \Hazy\ was incomplete. This description has now been amended. Note that in previous versions of \Hazy, the command was sometimes incorrectly called the \texttt{illuminate} command.

The \texttt{database H-like levels large} command now sets a minimum of 160 collapsed levels (was 10 in previous versions).

The \texttt{stop time <value>} command has been added to
allow integration of time-dependent simulations for a preset total amount of time, e.g., 20 Myr.
This functionality was used in a recent paper on the
mid-infrared emission in the Phoenix galaxy cluster
\citep{2025Natur.638..360R}.
It should also be useful to hydrodynamic simulations
that employ \Cloudy{} as a sub-grid process, or in
post-processing of simulation snapshots.

The following new commands have been added: \texttt{table SED available}, \texttt{abundances available}, and \texttt{grains available} with functionality similar to the \texttt{table star available} command. To enable the Ti-chemistry, the \texttt{set chemistry TiO on} command has been added. Also added were the following commands to monitor the behavior of the code: \texttt{monitor itrmax}, \texttt{monitor chemistry steps}, \texttt{monitor chemistry searches}, and \texttt{monitor time elapsed}.

The following commands have been removed: \texttt{set numerical derivatives}, \texttt{no fine opacities}, and \texttt{set H2 fraction}.

\subsection{Other changes}

\subsubsection{Additional Solar Abundance File}

\Cloudy{} includes a wide variety of solar system elemental abundance tables in its \textit{data/abundances/} directory, which have been compiled from the literature. Among these are the widely used solar abundance compilations from \citet{2009LanB...4B...44L} and \citet{2003ApJ...591.1220L}, both of which provide recommended values for a complete set of chemical composition of the solar system. For this particular release of \Cloudy{}, we have included a new file, \texttt{data/abundances/Lodders25.abn}. This file contains the latest solar abundance recommendations, as published by \citet{2025SSRv..221...23L}. This new dataset incorporates revised and updated solar photospheric abundances, which recovers a higher solar system metallicity.
This new abundance set is included
as an additional option to replace our
default solar composition, which is unchanged.

\subsubsection{Updated Fe II continuum bands}
In the file {\tt FeII\_bands.ini} it was stated that the lower and upper band edges would be treated as vacuum wavelengths. This was not quite how it worked as the vacuum band edges would be compared to air wavelengths of the lines in the standard setup. This has been fixed, resulting in changes in the predictions for all continuum bands. Especially the {\tt Fe 2b} 4971 and 7785 bands are strongly affected by this fix and a comparison with results from older \Cloudy\ versions is not meaningful.

The files {\tt FeII\_bands.ini} and {\tt continuum\_bands.ini} have been renamed to {\tt FeII\_bands.dat} and {\tt continuum\_bands.dat}, respectively, as they are not \Cloudy\ init files.
\subsubsection{New SED files}
We have added spectral energy distributions (SEDs) for NGC~5548 in both its obscured and unobscured states, based on the multiwavelength modeling presented by \citet{2015A&A...575A..22M}. These SEDs are now included in the \Cloudy{} data directory and can be used to model photoionized regions under realistic AGN conditions. The obscured SED represents the source during its 2013 absorption event, while the unobscured version corresponds to its historical, unobscured state.
Similarly, two new SEDs, obscured and unobscured, are added for Mrk~817. Both of these SEDs are explained and used in \citet{2021ApJ...922..151K, 2024ApJ...972..141D}

\subsubsection{New scripts}
To support ongoing STOUT updates and ensure compatibility with the latest NIST Atomic Spectra Database, we revised the \texttt{NistExtractor.py} script 
available in \texttt{cloudy/scripts/NistExtractor}. The updated version now 
interfaces with the current NIST API and retrieves up-to-date atomic data. It 
also supports a broader range of ion name formats (e.g., O\_III, o\_iii,
o\_3), which are correctly interpreted as O$^{2+}$. The script outputs a STOUT-
compatible directory structure, including \texttt{.nrg}, \texttt{.tp}, and 
\texttt{.coll} files. Since NIST does not provide collision strengths, the \texttt{.coll} file is left empty. 

A new script has been added to \texttt{cloudy/scripts/citation-plot} to retrieve data from NASA ADS and track \Cloudy{}'s citations by version and year. Running
this script requires a personal \texttt{ADS-API-TOKEN}. A version of the 
generated citation plot is updated weekly on \Cloudy{}'s wiki page\footnote{\url{gitlab.nublado.org/cloudy/cloudy/-/wikis/home}}.

\section{Cloudy on the Web}
\Cloudy{} is supported by a variety of online platforms that provide users with access to code, documentation, training resources, and published results. The development team maintains these web-based resources to ensure the community has the tools and information needed to run, understand, and properly cite \Cloudy{} simulations:
\begin{itemize}
   
 \item  \Cloudy{} is supported by a robust web presence that provides access to documentation, data, and source code. The official website, \texttt{nublado.org}, hosts installation guides, tutorials, atomic data descriptions, and links to recent \Cloudy{} releases. Users can also explore historical and current versions of the code and data through our GitLab repository, accessible from the website. \Cloudy{}'s main developers actively maintains a record of published versions on Zenodo, where users can obtain DOI-referenced software packages and associated datasets.

\item \Cloudy{}'s YouTube channel\footnote{\url{youtube.com/@Cloudy-Astroph}} provides tutorials and instructional videos designed to help users effectively run and interpret simulations with the \Cloudy{} spectral synthesis code. It covers a range of topics, from beginner introductions to advanced modeling techniques, and is regularly updated with new material, serving as a valuable learning resource for the \Cloudy{} user community.

\item \Cloudy{}'s Papers GitLab repository\footnote{\url{gitlab.nublado.org/cloudy/papers}} is a central location for accessing scripts used in our published papers related to the \Cloudy{} project. It includes associated figures and scripts used in the publications. This repository is maintained by the \Cloudy{} development team to ensure transparency, reproducibility, and accessibility of the scientific results.

\item \Cloudy{} has a long history of development, with multiple released versions and their corresponding documentation sets, known
as the Hazy manuals. These versions --- such as C90, C13, and
C17 --- are listed alongside their respective Hazy references in Table\,\ref{tab:CloudyReleasePapers}. While these prior versions 
remain archived for reproducibility and legacy support, the authors 
of \Cloudy{} strongly encourage users always to use the most recent version. This ensures access to the most up-to-date atomic data, 
physical processes, and bug fixes. Accordingly, users should cite 
the latest \Cloudy{} release and Hazy documentation to reflect the 
current state of the code and to support the ongoing development of the project. A complete citation for the current version of \Cloudy{} can be obtained by including the command \texttt{print citation} in
your input script when running the code. This will print the appropriate reference to cite in your output file.
\end{itemize}


\section{Future Directions}
\label{sec:future}

Since its inception in 1978, \Cloudy{} has undergone significant development with each new release. Its core mission has been to provide the astronomical community with a robust tool for interpreting the light emitted by astrophysical objects, in support of both space- and ground-based observatories. 
The field of astronomy is rapidly evolving, driven by upcoming missions capable of probing the earliest galaxies to the detailed atmospheric composition of exoplanets. Future observatories will offer improved sensitivity and resolution, surveying vast regions of the sky in unprecedented detail, and placing greater demands on the physical accuracy of simulation tools. 
Continued development of \Cloudy{} will therefore be essential to support both standalone applications of \Cloudy{} and its integration with next-generation hydrodynamic and machine learning codes. 
We outline below the two key areas in \Cloudy{} development needed to support future science:

\begin{itemize}
    \item State-of-the-art of general relativistic magnetohydrodynamic codes, are being developed to study multi-timescale, multi-wavelength, and multi-messenger astrophysical plasmas. These simulations depend critically on accurate atomic and molecular data within plasma environments, an area where \Cloudy{} plays a central role. To continue supporting these advanced hydrodynamic codes, \Cloudy{}'s atomic data must be expanded to provide high-fidelity atomic models.
    
    \item The search for potential life beyond solar-system has been rapidly growing over the past two decades, advancing from detecting individual extrasolar planets to detailed studies of exoplanet atmospheres and assessing their potential for hosting life. Current missions such as TESS and JWST are enabling insights to atmospheric chemistry, and upcoming missions such as the Roman Space Telescope and Habitable worlds observatory will directly image exoplanets in search for biosignatures. Major development must be undertaken to translate \Cloudy{}'s chemical network to simulating spectra from these exoplanet atmospheres.
\end{itemize}

\subsection{Atomic Data}
As part of our ongoing effort to improve \Cloudy{}'s atomic database, we plan to extend the same comprehensive framework used for the C-like isoelectronic sequence to other sequences. In particular, upcoming updates will focus on the Li-like, F-like, Ne-like, N-like, Mg-like, and O-like isoelectronic sequences within the Stout database. These enhancements will ensure consistency across ions and improve the accuracy of \Cloudy{}'s predictions across a wider range of astrophysical environments. These additions will also improve the physical treatment of high-lying levels, which plays a key role in mediating the transition to statistical equilibrium at high densities. These upper levels become significantly important as they approach the continuum, where the distinction between bound and free states becomes blurred. Properly modeling this regime is essential to capture continuum-lowering effects and to ensure thermodynamic consistency of the simulations.

\subsection{Molecular Data}

Currently, only 47 out of the 191 molecules included in \Cloudy{} have associated spectral lines. In the future, we plan to incorporate internal structures for the remaining molecules to enable the prediction of their spectral lines. In addition, we will include higher vibrational and rotational levels in the molecular models to better support the JWST observations. These enhancements will enable \Cloudy{} to accurately simulate non-local thermodynamic equilibrium (non-LTE) conditions- an essential capability for modeling hot exoplanets, where departures from LTE are common.

\subsection{Cloudy at high densities}
\Cloudy{}’s goal is to provide reliable results for densities ranging from the
low-density limit to densities where the system reaches
Local Thermodynamic Equilibrium (LTE) or Strict Thermodynamic Equilibrium (STE).
This is a challenging task due to the uncertain physics of highly-excited states.
Shown in Figs 10 and 11, the 2017 \Cloudy{} release paper \citep{2017RMxAA..53..385F} details that the models of one and two-electron systems are well behaved at all densities,
and level populations go to the proper thermodynamic limits at high densities. 
Figs 17 and 18 of the 2013 release paper
\citep{2013RMxAA..49..137F} show that the chemistry, 
ionization, and energy exchange go to the proper 
thermodynamic limits for a broad range of densities.
This 2025 \Cloudy{} release includes a major expansion
in the treatment of excited states with the
adoption of a large body of atomic data computed with the R-matrix suite of codes
\citep{2025Atoms..13...44D}.
This improves the physical treatment of the highest levels that mediate the approach to statistical equilibrium.

Substantial questions remain. The theory of continuum lowering at high densities
is the greatest uncertainty.
\citet{2022PASP..134g3001A}
discuss continuum lowering and its effects on the partition function.
Section 3 of that paper shows that the three available theories for continuum lowering
at high densities disagree by distressing amounts. A proper theory of
dense-plasma continuum lowering remains an unsolved grand challenge problem in physics.

Dielectronic recombination is often the dominant process for complex ions
\citep{AGN3}.
This occurs through highly excited and autoionizing states that are
greatly affected by continuum lowering.
Nigel Badnell and co-workers have created a theory for this suppression
and provided numerical fits to density-dependent dielectronic recombination
suppression factors 
\citep{2013ApJ...768...82N, 2018ApJS..237...41N, 2014JPhCS.488f2027G}.
These results are used by \Cloudy{} to account for high-density suppression of recombination.

The \Cloudy{} team participated in two of the ``NLTE\#'' series of meetings
\citep{2013HEDP....9..645C, 2017HEDP...23...38P}.
These compared predictions of codes designed for dense plasma laboratory experiments. 
Discussions at these workshops suggested that the leading cause for disagreement
between predictions of the various codes was the treatment of continuum lowering
upon dielectronic recombination.
This remains an uncertainty.

\subsection{Grain Depletions}

\citet{2022MNRAS.512.2310G} and \citet{2023MNRAS.520.4345G} introduced a revised elemental depletion scheme in \Cloudy{}, based on the work in \citet{2009ApJ...700.1299J}. 
This depletion framework provides a way for users to vary the overall elemental abundances depleted onto grains using a single parameter, $F_*$. However, \Cloudy{} currently treats the computation of depleted gas abundances independently from that of the elemental abundances locked into dust grains. 
Although these two components are intrinsically coupled, the code does not yet enforce a self-consistent depletion across both. 
Efforts are underway to address this limitation, in which the gas-phase and dust-phase abundances are computed in a mutually consistent manner, ensuring conservation of total elemental content.}

\acknowledgments
CMG and GF acknowledges support from NASA (19-ATP19-0188, 22-ADAP22-0139), JWST-AR-0628, JWST-AR-06419, and NSF (1910687). MD acknowledges support from STScI (JWST-AR-06419.001).
MC acknowledges support from NSF (1910687), NASA (19-ATP19-0188, 22-ADAP22-0139), 
and STScI (HST-AR-14556.001-A).
GS acknowledges the WOS-A grant from the Department of Science and Technology, India (SR/WOS-A/PM-2/2021).

\bibliography{Cloudy25.bib} 

\end{document}